\documentclass[10pt,prd,preprintnumbers,floatfix,aps,nofootinbib,showpacs,twocolumn,superscriptaddress,noshowpacs]{revtex4} 

\usepackage{epsfig}
\usepackage{url}
\usepackage{hyperref}

\usepackage[normalem]{ulem} 

\usepackage{latexsym}
\usepackage{epsfig}
\usepackage{amsmath}
\usepackage{amssymb}
\usepackage{wasysym}
\usepackage{graphicx}
\usepackage{verbatim}
\usepackage{enumerate,mdwlist}
\usepackage[titletoc]{appendix}
\usepackage{amsfonts}
\usepackage{tikz} 
\usetikzlibrary{calc}
\usepackage{pgfplots}
\usepackage[export]{adjustbox}

\usepackage{bbold}

\usepackage[normalem]{ulem}

\newcommand{\lp}{\left(}
\newcommand{\rp}{\right)}
\newcommand{\lb}{\left[}
\newcommand{\rb}{\right]}

\newcommand{\ba}{\begin{eqnarray}}
\newcommand{\ea}{\end{eqnarray}}
\newcommand{\be}{\begin{equation}}
\newcommand{\ee}{\end{equation}}

\newcommand{\cM}{c_{_M}}

\newcommand{\oDE}{\Omega_{_{DE}}}

\newcommand{\oDEo}{\Omega_{_{DE},0}}
\newcommand{\oMo}{\Omega_{_{M},0}}

\newcommand{\Mc}{\mathcal{M}_c} 
\newcommand{\Msun}{M_\odot} 
\newcommand{\R}{\mathcal{R}} 
\newcommand{\dL}{d_{_L}} 
\newcommand{\dLgw}{d_{_L}^\mathrm{gw}} 
\newcommand{\dLem}{d_{_L}^\mathrm{em}} 
\newcommand{\Ogw}{\Omega_{\mathrm{gw}}} 

\newcommand{\obs}{\mathrm{obs}}
\newcommand{\dd}{\mathrm{d}}
\newcommand{\detc}{\mathrm{det}}
\newcommand{\bbh}{\mathrm{bbh}}
\newcommand{\sgwb}{\mathrm{sgwb}}

\newcommand{\mmin}{m_\mathrm{min}}
\newcommand{\mmax}{m_\mathrm{max}}
\newcommand{\mbreak}{m_\mathrm{break}}
\newcommand{\mXX}{m_{99\%}}

\definecolor{grey}{rgb}{0.4,0.4,0.4}
\definecolor{dullmagenta}{rgb}{0.4,0,0.4}
\definecolor{darkblue}{rgb}{0,0,0.4}
\definecolor{midblue}{rgb}{0,0,0.5}
\definecolor{midred}{rgb}{0.5,0,0}
\definecolor{orange}{rgb}{1,0.5,0}
\definecolor{lightbrown}{rgb}{0.75,0.5,0.25}
\definecolor{tan}{cmyk}{0.14,0.42,0.56,0}
\definecolor{djunglegreen}{cmyk}{0.99,0,0.52,0}
\definecolor{lightgreen}{rgb}{0,1,0}
\definecolor{olivegreen}{cmyk}{0.64,0,0.95,0.40}
\definecolor{midgreen}{rgb}{0.0,0.675,0.0}
\definecolor{darkgreen}{rgb}{0,0.5,0}

\usepackage[all]{xy} 
\usepackage{amsfonts}

\begin{document} 

\title{Hearing gravity from the cosmos: \\ GWTC-2 probes general relativity at cosmological scales}

\author{Jose Mar\'ia Ezquiaga}
\email{ezquiaga@uchicago.edu; \\ NASA Einstein fellow}
\affiliation{Kavli Institute for Cosmological Physics and Enrico Fermi Institute, The University of Chicago, Chicago, IL 60637, USA}

\begin{abstract}
Gravitational-wave (GW) catalogs are rapidly increasing in number, allowing for robust statistical analyses of the population of compact binaries. Nonetheless, GW inference of cosmology has typically relied on additional electromagnetic counterparts or galaxy catalogs. I present a new probe of cosmological modifications of general relativity with GW data only. I focus on deviations of the GW luminosity distance constrained with the astrophysical population of binary black holes (BBHs). The three key observables are 1) the number of events as a function of luminosity distance, 2) the stochastic GW background of unresolved binaries and 3) the location of any feature in the source mass distribution, such as the pair instability supernova (PISN) gap. Despite a priori degeneracies between modified gravity and the unknown evolution of the merger rate and source masses, a large damping of the GW amplitude could be falsifiable since as redshift grows it reduces the events and lowers the edges of the PISN gap, which is against standard astrophysical expectations. Applying a hierarchical Bayesian analysis to the current LIGO--Virgo catalog (GWTC-2), the strongest constraints to date are placed on deviations from the GW luminosity distance, finding $c_{_M}=-3.2^{+3.4}_{-2.0}$ at $68\%$ C.L., which is $\sim10$ times better than multi-messenger GW170817 bounds. These modifications also affects the determination of the BBH masses, which is crucial to accommodate the high-mass binary GW190521 away from the PISN gap. In this analysis it is found that the maximum mass of $99\%$ of the population shifts to lower masses with increased uncertainty, $m_{99\%}=46.2^{+11.4}_{-9.1}M_\odot$  at $68\%$ C.L. Testing gravity at large scales with the population of BBHs will become increasingly relevant with future catalogs, providing an independent and self-contained test of the standard cosmological model. 
\end{abstract}

\date{\today}
 
\maketitle
 
\section{Introduction}

The first three observing runs of advanced LIGO \cite{TheLIGOScientific:2014jea} and Virgo \citep{TheVirgo:2014hva} have seen
a rapid growth in the number of gravitational wave (GW) detections \cite{LIGOScientific:2018mvr,Abbott:2020niy}  
indicating that the field will soon transition to the era of population analysis - where outliers will flag new phenomena, but the core science will arise from statistical analyses of many events. The current catalog of the LIGO--Virgo Collaboration (LVC) is known as GWTC-2.

Preparing in advance, population studies are already central to the LVC astrophysical program \cite{LIGOScientific:2018jsj,Abbott:2020gyp}. Among many interesting findings, GWTC-2 has shown support for the theory of pair instability supernova (PISN), which predicts a mass gap in the mass distribution of black holes 
between $\sim50-120\Msun$ \cite{Fishbach:2017zga,LIGOScientific:2018jsj,Abbott:2020gyp}. In their analysis only $2^{+3.4}_{-1.7}\%$ of binary black holes (BBH) 
have primary masses above $45\Msun$~\cite{Abbott:2020gyp}. 
Another key observable is the merger rate history \cite{Fishbach:2018edt}, which according to GWTC-2 
is probably growing with redshift, but not faster than the star formation rate  \cite{Abbott:2020gyp}. 

Although present astrophysical uncertainties play a crucial role in the interpretation of GW catalogs, population studies are not limited to modeling the source population. A good example are constraints on the cosmic expansion 
from the location of the lower and upper edge of the PISN gap \cite{Farr:2019twy,Ezquiaga:2020tns}. 
In general, mass distribution information allows to probe different background cosmologies \cite{Mastrogiovanni:2021wsd}. 
Beyond testing cosmological parameters, I will show that astrophysical population analyses can probe 
one of the pillars of the standard model of cosmology, namely, the validity of general relativity (GR) at large scales. 

Gravity can be tested with GW number counts \cite{Calabrese:2016bnu}, looking for deviations to the universal signal-to-noise ratio (SNR) distribution \cite{Schutz:2011tw,Chen:2014yla}. 
Nonetheless, such universal relation is only valid if the merger rate does not evolve with redshift. Therefore, this test only applies to the low-redshift universe \cite{Chen:2014yla}. 
However, cosmological modifications of the GW propagation are most relevant at high-redshifts since they accumulate over long travel distances. 
Some of these theories have been proposed to solve the $H_0$ tension \cite{Zumalacarregui:2020cjh,Abadi:2020hbr}. 
GW catalogs alone also probe waveform distortions \cite{Isi:2019asy,Abbott:2020jks}, 
GW lensing beyond GR  \cite{Ezquiaga:2020dao} and birefringence~\cite{Okounkova:2021xjv}.

Tests of gravity have seen a proliferation in light of multi-messenger GW astronomy \cite{Ezquiaga:2018btd}. If a prompt counterpart is detected, the speed of GWs can be constrained \cite{Lombriser:2015sxa,Bettoni:2016mij}, as beautifully exemplified with GW170817 \cite{Ezquiaga:2017ekz,Creminelli:2017sry,Baker:2017hug,Sakstein:2017xjx}. Moreover, by directly measuring the source redshift one can infer its electromagnetic (EM) luminosity distance (assuming a cosmology) and test differences w.r.t. the GW luminosity distance \cite{Belgacem:2017ihm,Belgacem:2018lbp,Belgacem:2019pkk,Mukherjee:2020mha}. 
After GW170817 \cite{Arai:2017hxj,Lagos:2019kds}, constraints on GR deviations were set to $\cM=-9^{+21}_{-28}$ at 68.3$\%$C.L \cite{Lagos:2019kds}, where $\cM=0$ defines GR (this parameter will be introduced later).\footnote{This multi-messenger constraint can be tighten if it is assumed that GW190521 had an associated counterpart \cite{Mastrogiovanni:2020mvm}, although current observations seem insufficient \cite{Ashton:2020kyr}.} 
Alternatively, one can use the GW localization volume to statistically infer the redshift with galaxy catalogs \cite{DelPozzo:2011yh}. 
Recent analyses of GWTC-2 find $\Xi_0=1.88^{+3.83}_{-1.10}$ at 68.3$\%$C.L for B-band and completeness threshold $P_\text{th}=0.2$ \cite{Finke:2021aom}, where GR is $\Xi_0=1$.
Despite the great promise of these multi-messenger tests, their applicability heavily relies on the number of bright multi-messenger mergers and the completeness of galaxy catalogs \cite{Finke:2021aom}. 
In contrast, this proposal only relies on GW data and can be considered as a guaranteed test. 

\section{BBH population and merger rates}

BBHs merge along the history of the universe following a comoving merger rate $\R(z)$. This quantity is highly model dependent and to present day mostly unknown. 
As a working hypothesis all BBHs will be assumed to be remnants of stars. 
Thus $\R(z)$ should be negligible at high redshift, before the peak of star formation $z_p$. 
To accommodate this astrophysical prior, 
I follow the parametrization 
\cite{Callister:2020arv}
\be \label{eq:Rz}
\R (z)=\R_0\, C_0\frac{(1+z)^\alpha}{1+\lp\frac{1+z}{1+z_p}\rp^{\alpha+\beta}}\,,
\ee
which peaks around $z_p$ and has a slope towards $z_p$ and after controlled by $\alpha$ and $\beta$ respectively.  
$C_0(z_p,\alpha,\beta)=1+(1+z_p)^{-\alpha-\beta}$ sets $\R(0)=\R_0$. 
Analyses of multiple populations together \cite{Ng:2020qpk} are left for future work.

To compute the number of detections one needs to include selection effects - how probable is to detect a binary with given intrinsic parameters. 
Following \cite{Abbott:2020gyp}, I take a broken power-law model for the primary mass $p(m_1)\propto m_1^{-\kappa_i}$ with sharp cutoffs at $\mmin$ and $\mmax$. The transition between the two slopes $\kappa_1$ and $\kappa_2$ occurs at a breakpoint $\mbreak$ (see App. \ref{app:statistical_formalism} for details). For the secondary mass I assume a uniform distribution between $m_\text{min}$ and $m_1$.  
Then, the selection effects will be encapsulated in the probability of detection $p_\text{det}$, which depends on the redshift and masses of the binary together with the detector network sensitivity. 
Altogether, the detection rate per redshift and component masses is
\be \label{eq:rate_z}
\frac{\dd^3 \dot{N}_\text{det}}{\dd z\dd m_1 \dd m_2}=\frac{\R(z)}{1+z}\frac{\dd V_c}{\dd z}p(m_1,m_2)p_\detc(z,m_1,m_2)\,,
\ee
where $V_c$ is the comoving volume. Spin priors follow \cite{Abbott:2020gyp}. 

Noticeably, if the source population and background cosmology are fixed, any modification in the number of events has to arise from the selection bias $p_\detc$. 
Precisely, modifications of gravity will change the SNR affecting the probability of detecting binaries at different redshifts. 

\section{Probing cosmological modifications of gravity}

Assuming that the emission and detection of GWs follows GR 
and that there are no additional tensor fields or chirality, beyond GR corrections can be encapsulated in the propagation equation 
\be
h'' + (2+\nu)\mathcal{H}h'+(c_g^2k^2+\Delta\omega^2)h=0\,,
\ee
for both polarizations  $h_{+,\times}$.  
Three possible modifications can occur: an anomalous propagation speed $c_g\neq c$, a modified dispersion relation $\Delta\omega^2\neq0$ and a change in the GW amplitude when $\nu\neq0$. 
Relevantly, $c_g$ has already been strongly constrained by GW170817 \cite{Monitor:2017mdv} and the modified dispersion relation can be probed directly searching for waveform distortions \cite{Abbott:2020jks}.\footnote{A modified dispersion could bias the parameter estimation, but given current constraints \cite{Abbott:2020jks} and number of events \cite{Moore:2021eok} it is reasonable to assume that they will not systematically affect the population inference in a dominant way.} 
For these reasons I concentrate on $\nu$ which uniquely determines (when $c_g=c$ \cite{Belgacem:2019pkk}) the relation between the GW luminosity distance $\dLgw$ and the EM luminosity distance $\dLem$:
\be \label{eq:dL_gw}
\frac{\dLgw(z)}{\dLem(z)}=\exp\lb\frac{1}{2}\int_0^z\frac{\nu(z')}{1+z'}dz'\rb\,.
\ee
In GR, $\left. \dLgw\right\vert_{\text GR}=\dLem=(1+z)\int_0^z\frac{c}{H(z)}dz$, where the last equality assumes flat cosmologies. 
The background cosmology is fixed to Planck2018 \cite{Aghanim:2018eyx}. 

Motivated by cosmological modifications of gravity which aim at explaining the present accelerated expansion, I will assume that the additional friction scales with the dark energy
\be
\nu(z)=\cM\,\frac{\oDE(z)}{\oDEo}\,,
\ee
where $\cM$ is a constant. Since the SNR scales inversely with the luminosity distance, $\rho/\rho_\mathrm{gr}=\dLem/\dLgw$, this leads to 
\be
\frac{\rho}{\rho_\mathrm{gr}}=\exp\lb\frac{-1}{2}\frac{\cM}{\oDEo}\log\lb\frac{1+z}{(\oMo(1+z)^3 + \oDEo)^{1/3}}\rb\rb\,,
\ee
which modifies $p_\detc$ in (\ref{eq:rate_z}). 
Of course, other parameterizations are possible \cite{Gleyzes:2017kpi}, being interesting to track directly $\dLgw/\dLem$ with the $(\Xi_0,n)$ model \cite{Belgacem:2018lbp}, 
but this is left for future work. 

Under these assumptions, the modifications of $\dLgw$ are described in very simple terms. If $\cM$ is positive, $\dLgw$ will be larger and the overall signal will be quieter. On the other hand, a negative $\cM$ reduces $\dLgw$ amplifying the GW. 
Similarly, the higher the redshift, the more important any of these effects become. 
Although modifications of $\dLgw$ are \emph{a priori} degenerate with arbitrary $\R(z)$ and $p(m_1,m_2)$, e.g. a louder signal could be interpreted as a closer or heavier source, beyond GR effects can lead to signatures that are against standard astrophysical expectations as I discuss next. 

\subsection{Detection rates}

\begin{figure}[t!]
\centering
\includegraphics[width = 0.99\columnwidth]{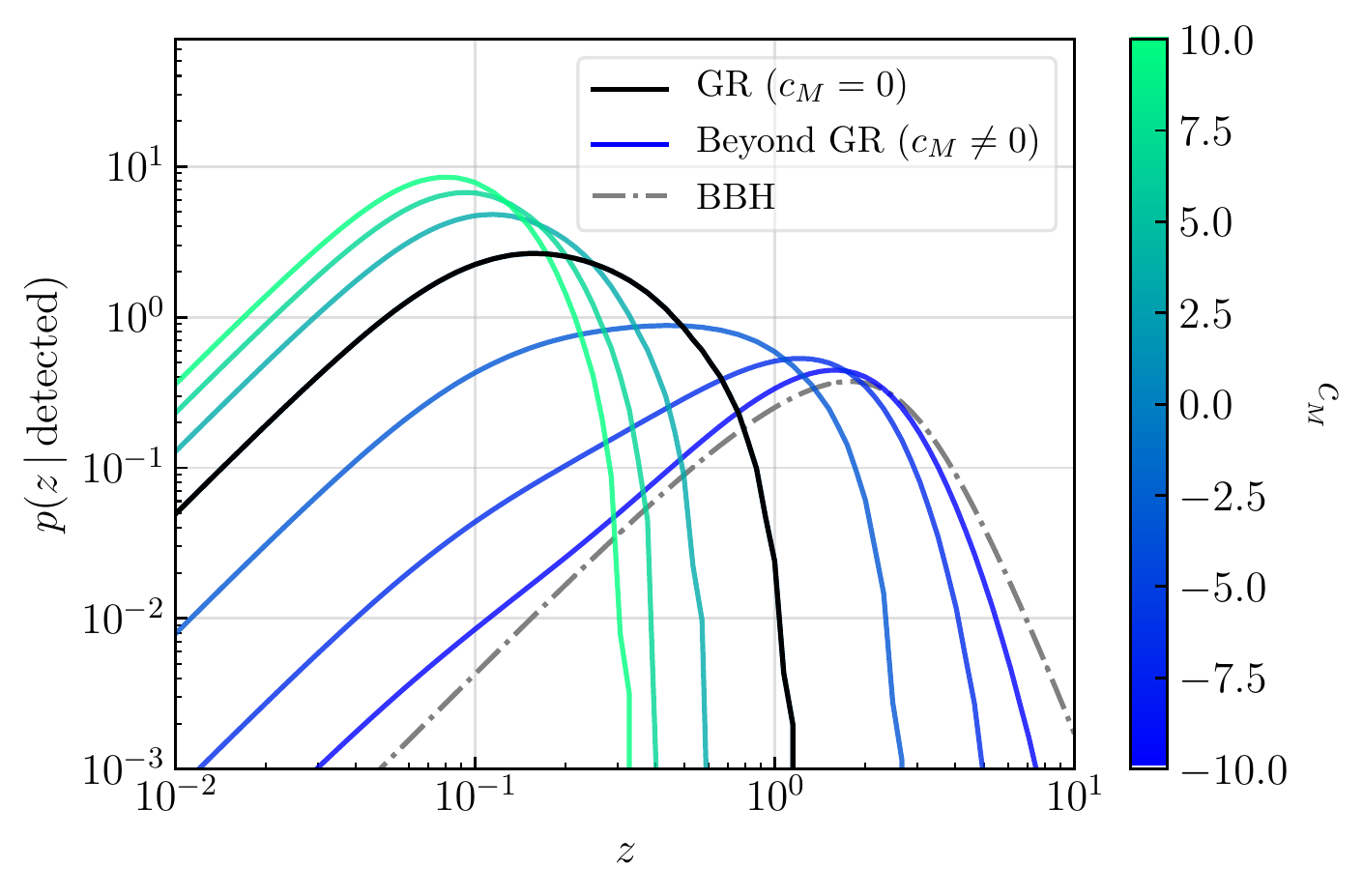}
 \caption{Observed redshift distribution of a population of BBHs within GR ($\cM=0$) and beyond ($\cM\neq0$) for LIGO/Virgo O3 sensitivity. 
The merger rate follows Eq. (\ref{eq:Rz}) with $\alpha=1.9$, $\beta=3.4$ and $z_p=2.4$; and the mass distribution a broken power law with $\mmin=5\Msun$, $\mmax=87\Msun$, $\kappa_1=1.6$, $\kappa_2=5.6$ and $b=0.43$.}
 \label{fig:pz_detected}
\end{figure}

If the cosmological propagation systematically changes the GW amplitude, it is easy to understand that this will affect the number of detections and how far one can hear them. 
This is explicit in Fig. \ref{fig:pz_detected} where the detected redshift distribution is plotted for present sensitivities. 
If $\cM \gg 1$, then only a fraction of the expected GR events are observed. On the opposite end, if $\cM\ll-1$ one detects events much further, eventually observing the entire population.  
This plot suggests that the shape of $p(z|\mathrm{detected})$ can have constraining power on modifications of $\dLgw$. Of course, any meaningful bound has to be placed allowing to vary the other unknown parameters of the model, as it will be done with the Bayesian inference.  

Despite the observed redshift distribution being correlated with the merger rate evolution, modification of $\dLgw$ can produce unexpected results by astrophysical priors breaking some of these degeneracies. 
For example, a decreasing rate of events with redshift before $z\sim1$ would conflict with BBHs following the star formation rate, which is known to increase up to $z\sim2$ \cite{Madau:2014bja}. 
This could serve to constrain $\cM\gg1$. 
Similarly, a modulation of the number of events with redshift would be astrophysically highly unexpected, but possible if GWs mix with other tensor fields introducing an oscillatory pattern in $\dLgw$ \cite{Jimenez:2019lrk}.
The latter nonetheless goes beyond the parametrization in Eq. (\ref{eq:dL_gw}). 

\subsection{Stochastic GW background of unresolved binaries}

Even though present detectors are only sensitive to relatively low-redshift events, the (non)-observation of the stochastic GW background (SGWB) produced by unresolved binaries provides valuable information. 
In fact, as shown in 
Fig. \ref{fig:pz_detected}, unless $\cM\ll-1$, only a small fraction of all mergers are being detected. 
The SGWB has the advantage that their sources are at higher redshift and thus more sensitive to modifications of $\dLgw$. 

The energy density of the SGWB can be computed summing over the energy flux emitted by all non-detected events, as determined by $1-p_\detc$. The dimensionless energy density $\Ogw$ scaling, including the inspiral phase only, is
\be \label{eq:Ogw}
\begin{split}
\Ogw(f)\sim f^{2/3}\int\int&\Mc^{5/6}\frac{\R(z)}{H(z)(1+z)^{1/3}}\lp\frac{\dLem}{\dLgw}\rp^2 \\
&\times p(\mathcal{M}_c)(1-p_\detc) \dd z\dd \Mc\,,
\end{split}
\ee
where the detailed derivation is deferred to appendix \ref{app:sgwb_mg} since it follows closely the classical result \cite{Phinney:2001di}. 
Interestingly, the ratio of luminosity distances appears quadratically in $\Ogw$. 
Although not included here, $\Ogw$ could constrain also modifications of the GW emission~\cite{Maselli:2016ekw,Saffer:2020xsw,Nunes:2020rmr}. 

Deviations in $\dLgw$ do not alter the typical $f^{2/3}$ spectral shape of $\Ogw$. 
However, they shift the turnaround point of the spectrum at $f>100$Hz. A positive $\cM$ moves the maximum to higher frequencies because the quieter sources behave as lighter ones reducing the effective minimum mass of the population. 
The peak of $\Ogw$ is, unfortunately, beyond ground-based detector sensitivities preventing the detection of this possible signature of modified gravity. 

\subsection{Source mass distribution}

\begin{figure}[t!]
\centering
\includegraphics[width = \columnwidth]{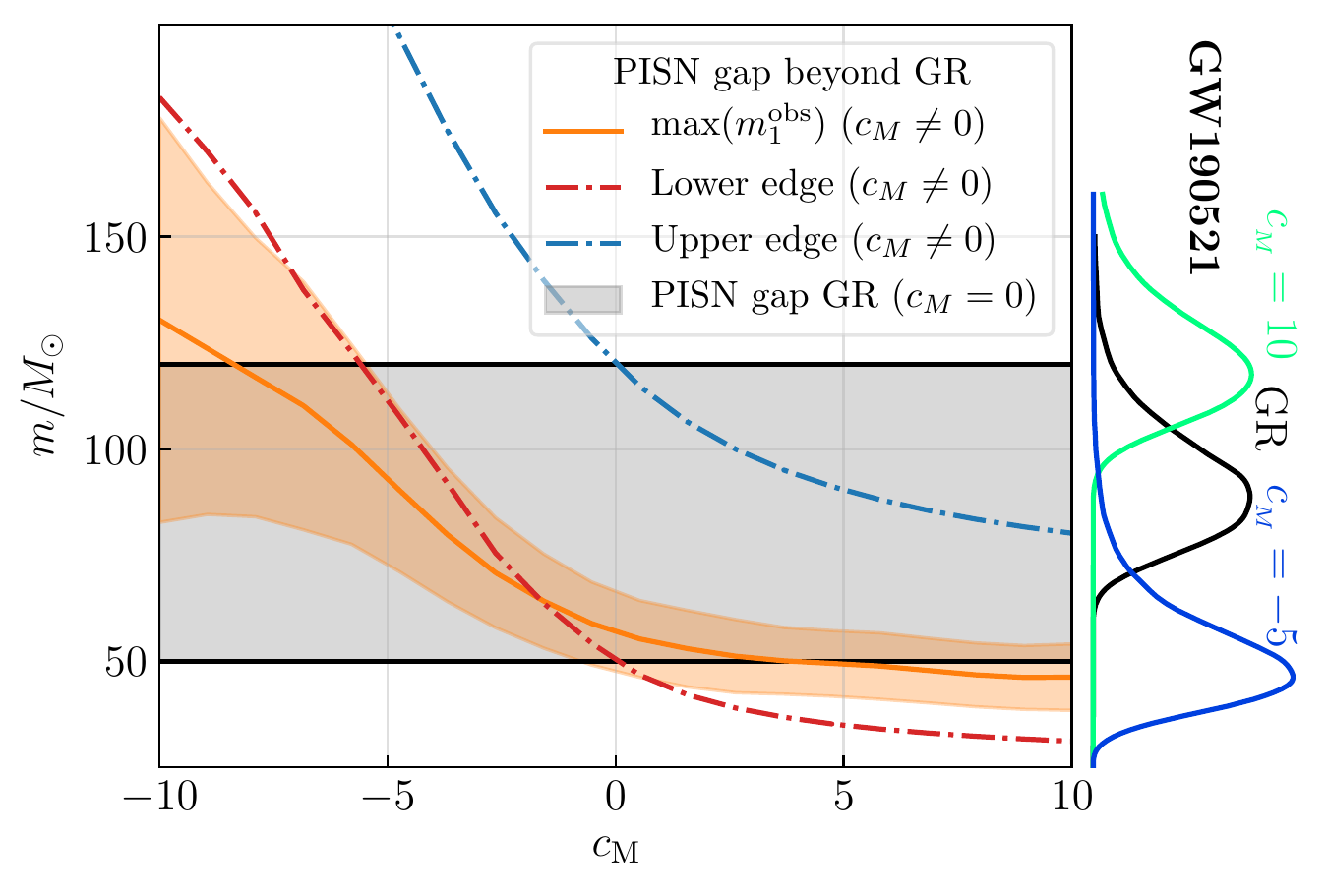}
 \caption{Impact of the modified luminosity distance on the inference of the PISN mass gap, fixed at $50-120\Msun$.  
 In the main panel the solid orange line represents the maximum inferred source frame mass with 39 detections  
 averaged over 100 mock simulations including selection bias for O3 sensitivity. The orange band indicates the $1\sigma$ dispersion. Dashed-dotted lines correspond to the displacement of the edges of the mass gap computed from the horizon distance in modified gravity. 
On the right, the posteriors for the primary mass of GW190521 \cite{Abbott:2020tfl} are presented for different values of $\cM$.}
 \label{fig:gap_inference}
\end{figure}

Modifying $\dLgw$ will bias the inferred source masses. 
This is particularly relevant when the distribution of masses $p(m_1,m_2)$ has a distinct mass scale, since this will break the degeneracy with the modified $\dLgw$. 
For BBHs, PISN theory sets two reference scales: the edges of the gap. 
A $\dLgw$ beyond GR will change the inferred location of the gap as exemplified in Fig.~\ref{fig:gap_inference}. 
Negative $\cM$ moves the PISN gap to higher values and vice-versa. 
This is because $\cM<0$ allows to expand the horizon redshift and apparently massive events could be just at higher redshift. 
In particular, the primary, source mass posterior of GW190521, the most massive event so far \cite{Abbott:2020tfl}, shifts according to the sign of $\cM$ as displayed in the right side of Fig. \ref{fig:gap_inference}. Negative $\cM$ can place GW190521 below the gap and large positive $\cM$ above it, in the ``far side" \cite{Ezquiaga:2020tns}.

These results are complementary to \cite{Straight:2020zke} where local modifications of gravity changing the PISN mass gap in the source population were studied. Here the modified propagation changes the inferred mass gap location, but not $p(m_1,m_2)$. 
In addition, this idea could be extended to other source populations. 
For instance, binary neutron stars (BNSs) have a narrow mass distribution 
that could constrain $\dLgw$ when tidal effects identify the compact object as a neutron star \cite{Messenger:2011gi}. 
Similarly, if the BNS merger rate was measured with EM observations, next generation GW detectors could tightly constrain $\cM$ since $\R(z)$ would be fixed \cite{Ye:2021klk}. 

Throughout this analysis the mass distribution is assumed to be constant in time, but 
modifications of $\dLgw$ would be roughly equivalent to change the source masses at different redshifts by $\tilde{m}_i(z)\approx (\dLem/\dLgw)^{6/5}m_i$. 
Thus, a $\cM>0$ emulates decreasing the maximum mass with redshift. 
This contradicts astrophysical expectations of $\mmax$ increasing with redshift due to the decrease in metallicity \cite{Dominik:2014yma}.   
Therefore, the evolution of the edges of the PISN gap could be a key determinant to test modifications of gravity. 
Recently, \cite{Fishbach:2021yvy} have shown evidence of an increase of $\mmax$ for a power-law model. 

\section{Constraints from GWTC-2} \label{sec:gwtc2}

To test cosmological modifications of gravity with current data, I develop a hierarchical Bayesian pipeline. Since this statistical framework is by now widely used in the GW community, details 
are presented in App. \ref{app:statistical_formalism}. 
The key differences with the standard analysis are:
\begin{enumerate}[i)]
\item source masses and redshifts inferred values depend on $\cM$; $m_i(m_{iz},\dLgw,\cM)$ and $z(\dLgw,\cM)$,
\item the probability of detection is also a function of $\cM$; $p_\detc(z,m_1,m_2,\cM)$.
\end{enumerate} 
With these considerations at hand, I analyze GWTC-2, using the same detection threshold as the LVC 
\cite{Abbott:2020gyp}. 
The parameters are: $\cM$ for the modification of gravity, $\{\R_0,\,\alpha,\,\beta,\,z_p\}$ for the merger rate history and $\{\kappa_1,\kappa_2,\mmax,b\}$ for the broken power-law mass distribution. The minimum mass of the population is fixed to $5\Msun$.  
Prior choices are specified in App. \ref{app:statistical_formalism}. 

\begin{figure}[t!]
\centering
\includegraphics[width = \columnwidth]{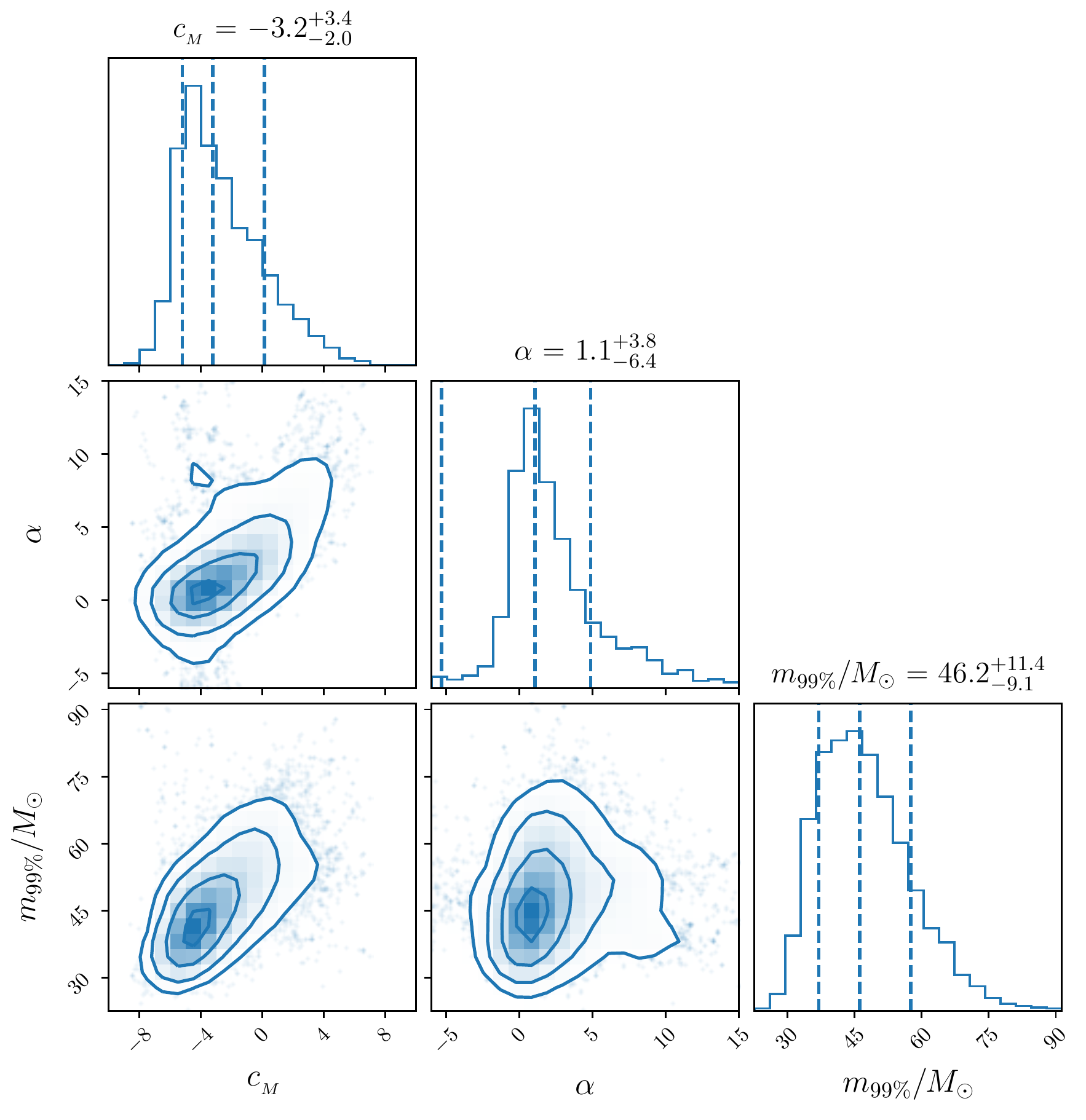}
 \caption{Posterior distributions for the modification of the GW luminosity distance $\cM$, the slope of the BBH merger rate $\alpha$ and the maximum mass of 99$\%$ of events $\mXX$ inferred from GWTC-2. Vertical lines correspond to the mean and $68\%$ confidence interval. 
Constraints on $\cM$ are $\sim 10$ times better than current multi-messenger bounds from GW170817~\cite{Lagos:2019kds}. 
Both $\alpha$ and $\mXX$ are correlated with $\cM$, with negative values of $\cM$ leaning towards the predictions of the theory of PISN.  
}
 \label{fig:summary_post}
\end{figure}

The main results are summarized in Fig. \ref{fig:summary_post} where the posterior distribution for $\cM$, $\alpha$ and $\mXX$ are presented. $\mXX$ corresponds to the maximum mass of $99\%$ of the events and can be derived directly from the posteriors of the mass distribution. 
There are several important results. 
First, the modification of gravity can be tightly constrained with BBH data only to $\cM=-3.2^{+3.4}_{-2.0}$ at $68\%$ C.L. This is $\sim 10$ times better than current multi-messenger constraints \cite{Lagos:2019kds}. 
Second, the merger rate slope is still likely positive ($\alpha>0$ at 65$\%$ probability) but its uncertainty increases w.r.t to the LVC results. 
This is due to the degeneracy between $\cM$ and $\alpha$. 
Finally, $\mXX$ shifts to smaller masses with larger errors, $\mXX=46.2^{+11.4}_{-9.1}\Msun$  at $68\%$ C.L., when compared to the LVC uniform-in-comoving-volume ($\alpha=\beta=0$) results ($\mXX=57.8^{+12.5}_{-8.7}$ at 90$\%$ C.L. \cite{Abbott:2020gyp}). Therefore, allowing for a modified GW propagation makes GWTC-2 
lean towards the theory of PISN. 

In addition, it is found that the current non-detection of the SGWB does not impose stronger constraints on $\cM$ than individual events. 
Moreover, GWTC-2 has not enough high-redshift sources to constrain $\beta$ and $z_p$ in the parametrization of $\R(z)$. Constraints on the other mass distribution parameters do not change significantly w.r.t. to the LVC results \cite{Abbott:2020gyp}. 
For completeness, full posterior samples are presented in Fig. \ref{fig:all_post} of App. \ref{app:posteriors}. 
It has been verified that the inferred model is consistent with the observed data performing a posterior predictive check analogous to \cite{Abbott:2020gyp}. 

Although the bounds on $\cM$ are subject to BBH population modeling, the parametrization chosen is flexible enough so that, under the astrophysical origin assumption, these results are robust. 
The fact that there is a preference for $\cM<0$ is consistent with astrophysical expectations that the PISN gap might increase with redshift \cite{Fishbach:2021yvy}. 
Bounds on $\cM>0$ are driven in part by the PISN-motivated prior on $\mmax\leq100\Msun$. 
An exploration of different population models/priors will be addressed in the future. 
When restricting to scalar-tensor theories, these GW constraints are comparable to present bounds from cosmological data \cite{Traykova:2019oyx}, although future survey expect to improve the latter by several orders of magnitude~\cite{Alonso:2016suf}. 
LISA standard sirens could complement LSS bounds \cite{Baker:2020apq}.

\section{Future prospects} \label{sec:conclusions}

GW observations contain a wealth of information about our universe. 
In this letter I have proposed a new probe of gravity at cosmological scales using the population of BBHs. 
This test requires GW data only and is thus a guaranteed output of any present or future GW catalog. 

Applying a hierarchical Bayesian analysis to GWTC-2, I find that current BBH observations constrain gravity more strongly than multi-messenger observations from GW170817, with overall results being consistent with GR. 
This is because modifications in the GW luminosity distance dramatically alter the inferred redshift and source mass distributions. In particular, deviations in $\dLgw$ w.r.t. GR shift characteristic scales in the source masses as the expected PISN mass gap. 
The effect of damping the GW amplitude with redshift ($\cM>0$) is particularly falsifiable since it leads to rates and PISN mass gap edges that decrease with redshift, which is against standard astrophysical predictions. 

This analysis can be extended to incorporate other parameterizations of $\dLgw$, being particularly interesting to test GW oscillations imprinting modulations in the observed redshift distribution \cite{Jimenez:2019lrk}. 
Similarly, including waveform distortions due to a modified dispersion relation could provide additional constraints beyond GR theories \cite{Mastrogiovanni:2020gua}. 
Although the background cosmology has been fixed throughout the analysis, BBHs observations can also be used to constrain $H_0$ and $\Omega_m$ \cite{Mastrogiovanni:2021wsd}.  A background and perturbation analysis would in principle be possible, but probably only for the scope of next generation detectors. 

This $\dLgw$ test can also be applied to other BBH populations, as those LISA will hear from space~\cite{Audley:2017drz}. From all LISA sources, the ones at higher redshift such as extreme-mass-ratio inspirals and super-massive black holes would be more interesting. 
Lensing effects could be incorporated in a similar way, with the probability of detection modified by the optical depth. 

BBH observations have proven to be a powerful test of gravity at cosmological scales. Future GW observations will only improve our understanding of the cosmological model.

\begin{acknowledgments}
I am grateful to the past and present University of Chicago LIGO group (Reed Essick, Amanda Farah, Maya Fishbach, Daniel Holz and Mike Zevin) for insightful conversation about GW population analyses, as well as the LIGO--Virgo R\&P and Cosmo groups for feedback on the results. I also acknowledge feedback on the manuscript from Maya Fishbach, Max Isi, Macarena Lagos, Simone Mastrogiovanni, Suvodip Mukherjee and Miguel Zumalac\'arregui. 
This analysis has used emcee \cite{ForemanMackey:2012ig} for the MCMC and corner \cite{corner} to present the posteriors. 
I am supported by NASA through the NASA Hubble Fellowship grant HST-HF2-51435.001-A awarded by the Space Telescope Science Institute, which is operated by the Association of Universities for Research in Astronomy, Inc., for NASA, under contract NAS5-26555. I am also supported by the Kavli Institute for Cosmological Physics through an endowment from the Kavli Foundation and its founder Fred Kavli. 
This research has made use of data, software and/or web tools obtained from the Gravitational Wave Open Science Center (https://www.gw-openscience.org/), a service of LIGO Laboratory, the LIGO Scientific Collaboration and the Virgo Collaboration.
\end{acknowledgments}

\appendix

\section{Statistical analysis}
\label{app:statistical_formalism}

In this appendix I summarize the hierarchical Bayesian pipeline developed in this analysis (see e.g. \cite{Mandel:2018mve} for a general discussion of this statistical framework). 
The first step is, of course, Bayes theorem. 
The posterior distribution of a given set of parameters $\Lambda$ describing a given population of BBHs follows from
\be
p(\Lambda|\{d_i\})\propto p(\{d_i\}|\Lambda)\pi(\Lambda)\,, 
\ee
where $p(\{d_i\}|\Lambda)$ 
is the likelihood of obtaining $N_\obs$ GW events with data $\{d_i\}$, while $\pi(\Lambda)$ are the prior expectations on $\Lambda$. 
Information about the stochastic background can also be included by the likelihood product
\be
p(\{d_i\},\Ogw|\Lambda)=p_\bbh(\{d_i\}|\Lambda)\times p_\sgwb(\Ogw|\Lambda)\,,
\ee
setting for example that the SNR of $\Ogw$ should be less than 2 during O3a. 
However this will not be included in the final results since it is found that it does not constrain more than individual events.

The likelihood of resolvable events can be described by a Poissonian process
\be
\begin{split}
p_\bbh(\{d_i\}|\Lambda)\propto &N_\detc(\Lambda)^{N_\obs}e^{-N_\detc(\Lambda)} \\
&\times\prod_{i=1}^{N_\obs}\frac{1}{\xi(\Lambda)}\left<\frac{p(\phi_i|\Lambda)}{\pi_\mathrm{pe}(\phi_i)}\right>_\mathrm{samples}\,,
\end{split}
\ee
where $\xi=N_\detc/N_\bbh$ is the ratio between the expected detected mergers $N_\detc$ and the actual merger $N_\bbh$. 
Note that the data likelihood, $p(d_i|\phi)$, given the GW parameters $\phi$, is not directly accessible. Instead there are only the event posteriors samples $p(\phi|d_i)$ to which it is necessary to factor out the prior used in the parameter estimation $\pi_\mathrm{pe}(\phi)$. 

The main observables are the inferred redshifts and source masses, thus $\phi=\{z,m_1,m_2\}$. Remember that, as explained in the main text, these three quantities depend on $\cM$ and are derived from the observed data of $\{m_{1z},m_{2z},\dLgw\}$. For these parameters the only relevant parameter estimation prior is $\pi(\dLgw)\propto(\dLgw)^2$ since for the masses the LVC uses a uniform prior. The prior $\pi_\text{pe}(z,m_1,m_2)$ is directly obtained including the Jacobian \cite{LIGOScientific:2018jsj}:
\be
\pi_\text{pe}(z,m_1,m_2)\propto (\dLgw)^2 (1+z)^2\frac{\partial\dLgw}{\partial z}\,,
\ee 
where in this case
\be
\frac{\partial\dLgw}{\partial z} = \frac{\dLgw}{1+z}+\frac{(1+z)c}{H(z)}\frac{\dLgw}{\dLem}+\frac{\nu}{2(1+z)}\dLgw\,, 
\ee 
following Eq. (\ref{eq:dL_gw}). 
Altogether, the BBH likelihood can be written as 
\be
\begin{split}
p_\bbh(\{d_i\}|\Lambda)\propto &\,e^{-N_\detc(\Lambda)} \times\prod_{i=1}^{N_\obs}\left<\frac{dN(\phi|\Lambda)/d\phi}{\pi_\mathrm{pe}(\phi)}\right>_\mathrm{samples}\,.
\end{split}
\ee
This result can be simplified further if the local merger rate $\R_0$ is marginalized using a uniform in log prior to obtain
\be
\begin{split}
p_\bbh(\{d_i\}|\Lambda)\propto &\,\xi^{-N_\obs} \times\prod_{i=1}^{N_\obs}\left<\frac{p(\phi|\Lambda)}{\pi_\mathrm{pe}(\phi)}\right>_\mathrm{samples}\,,
\end{split}
\ee
which does not depend on $\R_0$.

The BBH population is modeled with a merger rate history following Eq. (\ref{eq:Rz}). 
For the primary mass a broken power-law distribution is used:
\begin{equation}
p(m_1)\propto
\begin{cases}
m_1^{\kappa_1}, \quad \mmin<m_1<\mbreak \\
m_1^{\kappa_2}, \quad \mbreak<m_1<\mmax \\
0, \quad \text{elsewhere}
\end{cases}\,,
\end{equation}
where $\mbreak = \mmin + b(\mmax-\mmin)$ and $b\subset(0,1]$. In the limit of $b\to1$ one finds $\mbreak\to\mmax$. 
On the other hand, the secondary source mass is uniformly sampled below $m_1$ and above $\mmin$. 
In the analysis the minimum mass is fixed to $5\Msun$. 
Therefore, in total the BBH population is modeled by 8 parameters:
$\Lambda_\bbh=\{\R_0,\alpha,\beta,z_p,\kappa_1,\kappa_2,\mmax,b\}$. 
The modification of gravity is modeled with 1 parameter: $\Lambda_\mathrm{gravity}=\{\cM\}$. 
The priors are chosen to be uniform distributions in the ranges: $\log_{10}\R_0\subset[-3,3]$, $\alpha\subset[-25,25]$, $\beta\subset[0,10]$, $z_p\subset[0,4]$, $\kappa_{1,2}\subset[-4,12]$, $\mmax\subset[30,100]$, $b\subset[0,1]$ and $\cM\subset[-12,12]$. 
The probability of detection during O3a is computed using the public sensitivity of matched filter searches.\footnote{https://dcc.ligo.org/LIGO-P2000217/public}

\section{Stochastic background of GWs with modified propagation}
\label{app:sgwb_mg}

In this appendix I provide a derivation of how the modified GW propagation affects the stochastic background of unresolved binaries. 
I focus in particular on modification in the GW luminosity distance. 
This derivation extends the classical result of \cite{Phinney:2001di} beyond GR, allowing for $\dLgw\neq\dLem$. 

The dimensionless stochastic GW background is defined as 
\be
\Ogw(f)=\frac{1}{\rho_c}\frac{\dd \rho_\text{gw}}{d\ln f}=\frac{f}{c\rho_c}F(f)\,,
\ee
where $\rho_c=3c^2H_0^2/8\pi G$ is the critical energy density and frequencies are in the detector frame. In the second equality, the total energy flux $F(f)=c\dd\rho_\text{gw}/\dd f$ is introduced. 
The total flux is nothing but the energy emitted by all binaries per unit area:
\be
F(f)=c\dot{N}\,\frac{\dd E_\text{gw}(f)}{d f}\frac{(1+z)^2}{4\pi(\dLgw)^2}\,.
\ee
The number of events per detector frame time has already been defined in Eq. (\ref{eq:rate_z}). There, in order to account for all the binaries which cannot be detected individually one simply needs to substitute $p_\detc\to (1-p_\detc)$. One obtains
\be
\frac{\dd^3 \dot{N}_\text{unresolv}}{\dd z\dd m_1 \dd m_2}= \frac{\R(z)}{(1+z)}\frac{\dd V_c}{\dd z}p(m_1,m_2)(1-p_\detc)\,.
\ee
The energy emitted per frequency is given by (recall only modification in the GW propagation are being considered)
\be
\frac{\dd E_\text{gw}}{\dd f} = (1+z)\frac{(G\pi)^{2/3}}{3}\Mc^{5/3}f_{s}^{-1/3}
\ee
during the inspiral of a circular binary. A more general expression can be obtained simply noting that $\dd E_\text{gw}/\dd f \sim \dL^2 f^2\langle| \tilde{h}(f) |^2 \rangle_{\vec{\Omega}}$ where $\tilde{h}(f)$ is the Fourier transform of the time domain strain (during the inspiral $\tilde h(f)\sim f^{-7/3}$) which has been averaged over all possible sky locations and orientations $\vec{\Omega}$. 
Noticeably, the number of events scales with the differential comoving volume 
\be
\frac{\dd V_c}{\dd z}=\frac{4\pi r(z)^2}{H(z)}=\frac{4\pi (\dLem)^2}{(1+z)^2H(z)}\,.
\ee
Importantly, the comoving rate scales with $(\dLem)^2$ while the GW energy flux scales with $1/(\dLgw)^2$. In GR these two quantities are equal and cancel each other. However, beyond GR they do not and $\Ogw$ depends on their ratio square $(\dLem/\dLgw)^2$. 
The final result for the inspiral signal is given in Eq. (\ref{eq:Ogw}). If one wants to include the full emitted signal this can be generalized to 
\be \label{eq:Ogw_general}
\begin{split}
\Ogw(f)= \frac{4\pi^2f^3}{3H_0^2}\int&|\tilde{h}_\text{gr}(f,\vec{\phi})|^2\frac{\R(z)}{(1+z)}\frac{\dd V_c}{\dd z}\lp\frac{\dLem}{\dLgw}\rp^2 \\
&\times p(m_1,m_2)(1-p_\detc(\vec{\phi})) \dd \vec{\phi}\,,
\end{split}
\ee
where $\vec{\phi}=\{m_1,m_2,z\}$ and $\dd\vec\phi=\dd z\dd m_1 \dd m_2$. Here $\tilde h_\text{gr}$ is the GR emitted signal (which is inversely proportional to $\dLem$ and thus this distance factor cancels with the one from $\dd V_c / \dd z$). 

\newpage

\section{Full posterior samples}
\label{app:posteriors}

\begin{figure*}[t!]
\centering
\includegraphics[width = \textwidth]{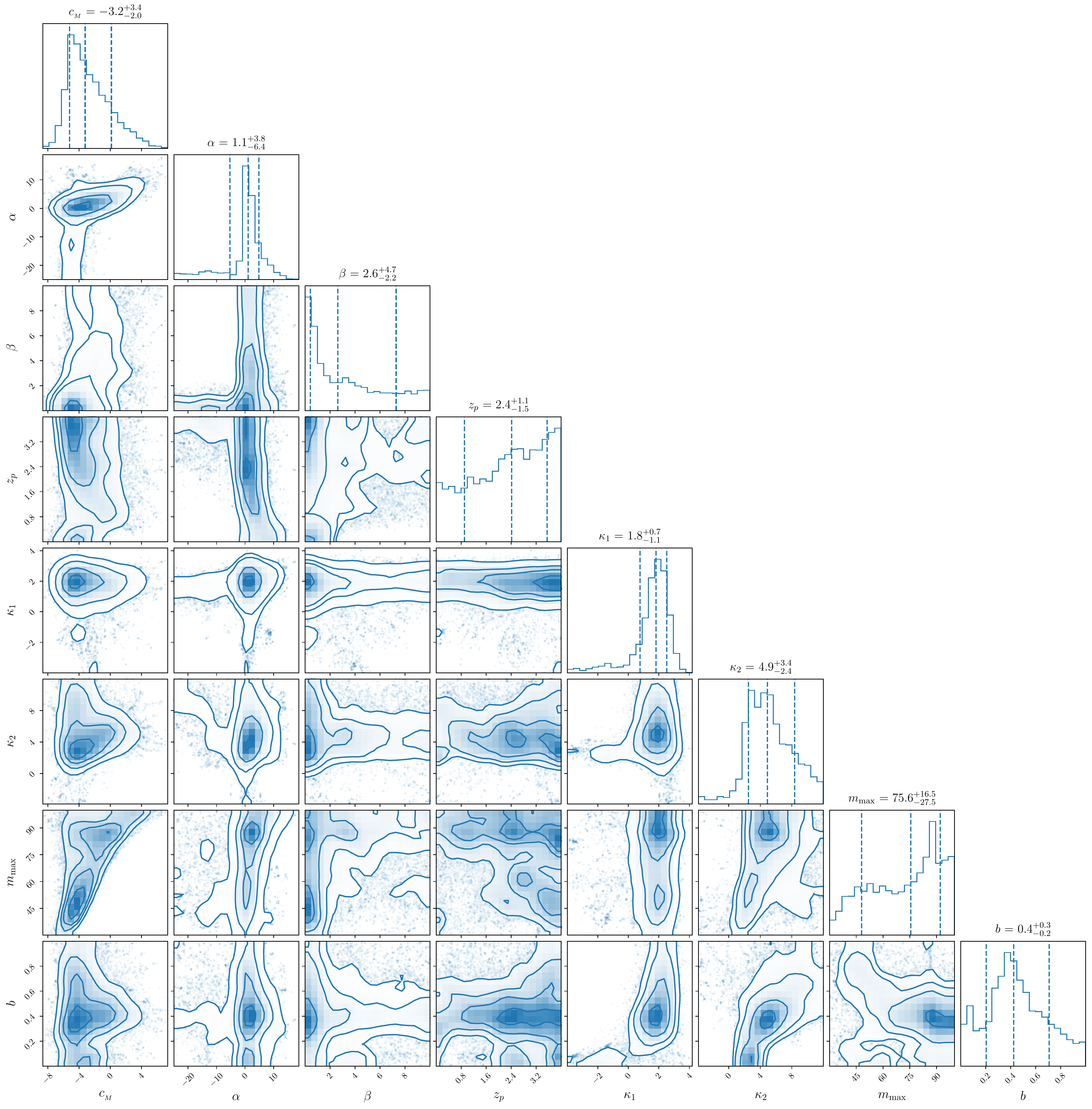}
 \caption{Posterior distributions from the analysis of modifications of the GW luminosity distance in the BBH population from GWTC-2. 
 The parameters of the analysis are $\cM$ for the modification of gravity, $\{\alpha,\,\beta,\,z_p\}$ for the merger rate history and $\{\kappa_1,\kappa_2,\mmax,b\}$ for the broken power-law mass distribution. 
 The local merger rate $\R_0$ has been marginalized using a uniform in log prior.}
 \label{fig:all_post}
\end{figure*}

For completeness I present in this appendix the full posterior distributions for all the parameters in the analysis. The results are displayed in Fig. \ref{fig:all_post}. 
It is to be noted that in Fig. \ref{fig:summary_post} the range in $\alpha$ was cut below $-6$. This is because as shown in this figure, the parametrization used in this analysis saturates at $\alpha\lesssim-5$ and the inference is the same. In any case, $85\%$ of the posterior is above this value.

\newpage

\bibliography{gw_refs}

\end{document}